\documentclass[aps,reprint,nofootinbib,superscriptaddress,amsmath,amssymb,longbibliography]{revtex4-1}

\usepackage{graphicx}
\usepackage{amsmath}
\usepackage{bm}
\usepackage{float}
\usepackage{hyperref}
\usepackage{xcolor}
\usepackage[T1]{fontenc}

\begin{document}

%Title of paper
\title{Towards principles of brain network organization and function}

\author{Suman Kulkarni}
\email{sumank@sas.upenn.edu}
\affiliation{Department of Physics \& Astronomy, College of Arts \& Sciences, University of Pennsylvania, Philadelphia, PA 19104, USA}
\author{Dani S. Bassett}
\email{dsb@seas.upenn.edu}
\affiliation{Department of Physics \& Astronomy, College of Arts \& Sciences, University of Pennsylvania, Philadelphia, PA 19104, USA}
\affiliation{Department of Bioengineering, School of Engineering \& Applied Science, University of Pennsylvania, Philadelphia, PA 19104, USA}
\affiliation{Department of Electrical \& Systems Engineering, School of Engineering \& Applied Science, University of Pennsylvania, Philadelphia, PA 19104, USA}
\affiliation{Department of Neurology, Perelman School of Medicine, University of Pennsylvania, Philadelphia, PA 19104, USA}
\affiliation{Department of Psychiatry, Perelman School of Medicine, University of Pennsylvania, Philadelphia, PA 19104, USA}
\affiliation{Santa Fe Institute, Santa Fe, NM 87501, USA}
\date{\today}

\begin{abstract} % Limit: 150 words 
The brain is immensely complex, with diverse components and dynamic interactions building upon one another to orchestrate a wide range of functions and behaviors. Understanding patterns of these complex interactions and how they are coordinated to support collective neural activity and function is critical for parsing human and animal behavior, treating mental illness, and developing artificial intelligence. Rapid experimental advances in imaging, recording, and perturbing neural systems across various species now provide opportunities and challenges to distill underlying principles of brain organization and function. Here, we take stock of recent progresses and review methods used in the statistical analysis of brain networks, drawing from fields of statistical physics, network theory and information theory. Our discussion is organized by scale, starting with models of individual neurons and extending to large-scale networks mapped across brain regions. We then examine the organizing principles and constraints that shape the biological structure and function of neural circuits. Finally, we describe current opportunities aimed at improving models in light of recent developments and at bridging across scales to contribute to a better understanding of brain networks.
\end{abstract}

\maketitle

\section{Introduction}
Recent years have witnessed phenomenal experimental strides in producing detailed large-scale maps of neural systems, as well as in probing and perturbing neural activity. Advances in electron microscopy and volumetric reconstructions now enable researchers to examine increasingly larger organisms at the cellular level to map out synaptic connections between neurons. While the synaptic connectivity of \emph{C. elegans} was first mapped in 1986 \cite{white1986structure}, we now have wiring maps of an adult fly brain \cite{dorkenwald2023neuronal,scheffer2020connectome}, portions of the mouse \cite{microns2021functional} and human cerebral cortex \cite{shapson2024petavoxel}, among other species \cite{ryan2016cns,veraszto2020whole,hildebrand2017whole}. These fine-scale maps add to approaches that probe connectivity at coarser scales, such as DTI and MRI \cite{li2013mapping,elam2021human,oh2014mesoscale}. In addition to these structural measures, technological advances in large-scale recordings of neuronal activity---such as optical calcium imaging and electrophysiological probes---now make it possible to measure the activity of thousands of neurons simultaneously \cite{urai2022large,weisenburger2018guide,steinmetz2021neuropixels,paulk2022large}. In fact, with Neuropixels probes, researchers can track neuronal activity at high spatio-temporal resolution in freely moving animals \cite{steinmetz2021neuropixels,paulk2022large,jun2017fully}. With this increasing access to large quantities of experimental data comes opportunities to develop quantitative methods that can cope with these large datasets and distill from them underlying principles of brain structure and function.

A central theme in the study of neural systems is to examine how macroscopic functions and behaviors arise from fine-scale interactions between neural elements. This perspective aligns well with the framework of statistical physics, which studies the macroscopic behavior of large ensembles of interacting microscopic entities. However, a major caveat here is that in traditional applications of statistical physics, the elementary components of studied systems are relatively simple and well understood. That is, collective effects result from the interaction of many `simple' elements rather than from the complex behaviors of the entities themselves. In contrast, the brain is extremely complex, characterized by heterogeneous and dynamic elements and interactions that span multiple scales. Still, despite this complexity and inherent messiness, one hopes to take some inspiration from statistical physics and attempt to build simple, tractable models that can capture important properties of the system. Here, we discuss opportunities that recent large-scale experimental efforts provide to glean principles of neural system design and function. Our account is explicitly multi scale, focusing on methods rooted in statistical physics, network theory and information theory.  

\begin{figure*}
    \centering
    \includegraphics[width=\linewidth]{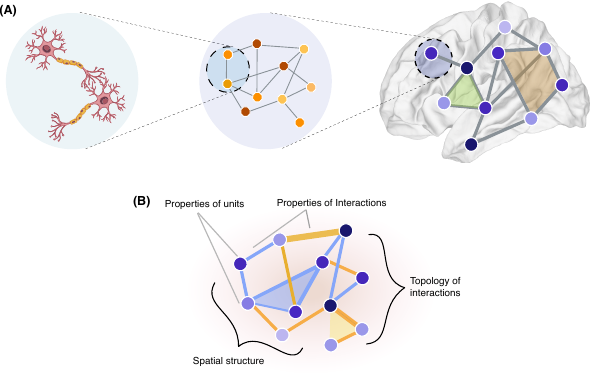}
    \caption{\textbf{Organization of the brain viewed at different scales.} (A) At the fine scale, neural systems are comprised of interacting neurons (\emph{left}). To study the activity arising out of collective behavior of groups of neurons, researchers build coarse-grained models describing networks of neurons (\emph{center}). To study the large-scale activity of an entire brain, researchers consider even coarser descriptions where nodes correspond to brain regions or neuronal populations (\emph{right}). (B) The activity and functioning of neural systems is governed by the individual properties of the units, the properties of their interactions, and the structure of interactions. Together, all these factors shape the dynamics \emph{of} the network itself and the dynamics of activity that occurs \emph{on} the network.}
    % I can include a tiny subset illustrating a feedback look between dynamics on and of networks?
    \label{fig:Fig1}
\end{figure*}

We structure our initial discussion in Sections II--IV based on spatial scales (as depicted in Fig. \ref{fig:Fig1}), focusing on the organization and modeling approaches relevant at each scale. It is important to note, however, that there is no clean separation between these scales and they are interdependent. We begin at the fine scale in Section \ref{sec:Sec2}, briefly describing detailed biophysical models of individual neurons to highlight features often not captured in coarser-scale models. In Section \ref{sec:Sec3}, we describe small populations of neurons, surveying modeling approaches and trade-offs involved. Section \ref{sec:Sec4} expands to large-scale brain networks, detailing methods to probe the structure and activity of large brain regions at both coarse and fine scales, and summarizing how tools from physics and mathematics are used to analyze these datasets. Having built this multi-scale picture in Sections II--IV, Section \ref{sec:Sec5} explores biological design principles and constraints that shape the properties and functioning of brain networks discussed throughout the paper.
In Section \ref{sec:Sec6}, we describe several critical frontiers: expanding theory and modeling to make sense of the latest connectome data, enhancing the feedback loop between theory and experiment, and using perturbative approaches to better understand neural systems. We conclude with an overview of formal accounts of explanation and causation in the context of neurophysics, underscoring the importance of contextualizing ongoing efforts.

\section{Organization and models at the cellular level}\label{sec:Sec2}

Though this review primarily focuses on large brain networks, it is crucial to discuss the neurons that make up the brain (see Fig. \ref{fig:Fig1}). Neuron biophysical properties affect both information processing within cells and collective behavior networks, such as synchronization \cite{hesse2022temperature,gowers2024neuronal}. Since synchronization is considered vital for information processing and altered neural synchronization is associated with brain disorders \cite{budzinski2019phase,uhlhaas2010abnormal,jiruska2013synchronization}, these properties have implication for health and disease. Morphologically, neurons are specialized structures that allow for communication and computations: dendrites branch out extensively to collect signals from other neurons, while the axon extends to transmit signals onwards. There is enormous diversity in the morphological, physiological and connectional features of neurons, which influences their functional properties \cite{lim2018development,masland2001neuronal,peng2021morphological}. Neurons transmit information in the form of electrical pulses (called spikes or action potentials) characterized by a rapid change in membrane potential over very short timescales, though there are classes of non-spiking neurons which generate graded potentials instead \footnote{including many neurons in the \emph{C. elegans} nervous sytem.} \cite{roberts1981neurones, juusola1996information,naudin2022simple, naudin2022systematic}. Once a spike is generated, neurons relay signals to neighboring neurons through synaptic connections that may be chemical (orchestrated through neurotransmitters) or electrical (through gap junction channels that connect neurons). Ultimately, a combination of all the excitatory or inhibitory influences on a neuron determine whether it generates an action potential. 

An cornerstone for computational neuroscience was the Hodgkin-Huxley model \cite{hodgkinhuxley1952}, developed in the 1950s to describe the generation of action potentials in a space-clamped squid giant axon. This detailed biophysical model describes the generation of action potentials through fast depolarizing and slow hyperpolarizing currents. The original model consists of a system of four coupled, nonlinear differential equations with parameters obtained from neurophysiological recordings. This model has been extended in several ways to study different kinds of neurons, for example incorporating additional ion channels \cite{hille2001ion} (as most neurons have many more ion channels) and stochastic elements \cite{goldwyn2011and}. At the same time, the relatively high dimensionality of the Hodgkin-Huxley model motivated simpler lower-dimensional models that can still reproduce many key features of neuronal dynamics such as the FitzHugh-Nagumo model \cite{fitzhugh1961impulses} and the Morris-Lecar model \cite{morris1981voltage,rinzel1998analysis}. These simplified models are far more amenable to mathematical analysis and facilitate efficient large-scale simulations of groups of neurons. A distinct approach is to consider phenomenological models that do not directly capture the detailed biophysics of spike generation. Examples of this include the integrate-and-fire model \footnote{which was originally introduced long before the mechanisms of spike generation were even known \cite{lapicque1907recherches}.}. At its heart, this model uses a differential equation to describe the membrane potential dynamics, and spikes are said to occur when the potential crosses a threshold, following which the potential is algorithmically reset to a certain value below the threshold. Several versions of this model exist with varying levels of biophysical realism, including nonlinear \cite{fourcaud2003spike} and adaptive integrate-and-fire models \cite{brette2005adaptive}. Another popular type of model is the Izhikevich model which combines some of the biologically realistic aspects of Hodgkin-Huxley type models with the computational efficiency of integrate-and-fire type models \cite{izhikevich2003simple}.

% Binary and artificial neuron models

As noted earlier, neurons are not point-like objects and have intricate spatial structure. Much of the total membrane area of neurons is occupied by the dendritic tree, whose structure and extent relates to the properties and firing rates of the neurons that it makes up \cite{london2005dendritic}. Dendrites are typically modelled using a cable equation describing the potential as a function of time and space, and the numerous branches can be accounted for using compartmental models. While ion channels can also be added to these compartments for more biological realism, a more tractable approach is to lump all processes responsible for spike generation in the soma and treat the dendrites as passive cables. In addition to dendrites, the length and the level of myelination of axons is known to affect signal conduction \cite{gerstner2014neuronal}. Evidently, there is astounding diversity in neuronal properties and morphology that can influence activity. Consequently, it is useful to classify action potentials into excitability classes according to the type of bifurcation that occurs close to the threshold \cite{rinzel1998analysis,ermentrout1996type,izhikevich2007dynamical}. The qualitative features of action potentials depend on the type of bifurcation. For more detailed descriptions of the models discussed in this section, we refer the reader to Refs.\cite{ermentrout2010foundations,gerstner2014neuronal,koch2004biophysics,rabinovich2006dynamical}

\section{Modelling small groups of neurons}\label{sec:Sec3}

In principle, one could keep adding detail to neuron models accounting for various biological nuances. However, there is a trade-off here between building models that are extremely detailed and building models that are tractable. This is particularly important for research questions aimed at understanding how large populations of neurons orchestrate activity, function and behavior. The challenge with detailed models is not solely a matter of the computational power required. These models rely on numerous biophysical parameters that can be difficult to measure reliably in experiments. Ideally, parameters for each neuron could be estimated from \emph{in vivo} experiments \cite{van2007neurofitter,willms1999improved,kim2019neural}. However, this requires experiments to collect data for each type of ion channel, many of whose details may still be unknown. When such experiments are infeasible, parameters are often tuned in an ad-hoc manner. Even when measured, these parameters are often obtained under specific conditions, making them less generalizable. Models with hundreds of parameters can also become difficult to reliably estimate, interpret and analyze. Even for a small and relatively well-studied organism like \emph{C. elegans}, which has 302 neurons, including the dozens of (known) ion channels can result in over a hundred parameters. Further, many parameter combinations can produce similar neural activity and network behavior \cite{prinz2004similar,marder2011multiple}. While this degeneracy may be a feature  rather than a drawback (possibly underpinning resilience)\cite{edelman2001degeneracy,prinz2004similar}, it makes detailed bottom-up approaches challenging.

Though detailed biophysical models are useful for studying cellular processes, it remains unclear how these detailed processes combine to generate the range of complex macroscopic behaviors observed, even in relatively simple organisms. Not all of these details may be relevant to higher-level processes in organisms. Moreover, experiments in humans and other large animals typically track the activity of regions, not individual neuron spikes. Arguments such as these align well with the principles of statistical physics, which provides a framework to study of macroscopic phenomena arising from large ensembles of interacting microscopic entities. To examine collective behavior, it often suffices to use coarse-grained descriptions instead of capturing every detail. Across fields of science, from physics to systems biology, several multi-parameter models display behaviors that largely depend only on a few combinations of parameters, with many other parameters being relatively unimportant for model behavior and prediction \cite{transtrum2015perspective, machta2013parameter}. Indeed, to build a theory of how sound waves propagate, we only need information such as the density and compressibility of matter, not the precise shape of each individual molecule. The goal is to build minimal models that reproduce the key features of interest. This approach is not to devalue details or lower-level explanations. Rather, the choice of model depends on the phenomena being studied and on the research questions guiding the investigation. Motivated by statistical physics, then, it is of interest to ask: Can we apply similar principles to the nervous system? How can we bridge fine-scale descriptions of brain organization with the functioning of nervous systems and the behaviors of organisms?

To model larger populations of neurons, researchers typically track the average activity of groups of neurons rather than modeling individual spikes of every neuron. Indeed, there is evidence that the brain tends to be organized into assemblies of neurons that display strongly correlated activity, operating as a functionally homogeneous unit \cite{breakspear2002nonlinear,deco2008dynamic}. Broadly speaking, there are two different approaches in constructing such lower-dimensional average-rate models. 
The first is to build phenomenological neural-mass models that capture observations at the scale of neuron groups. The second, more bottom-up approach involves developing approximations of spiking neuron models. Such population-level models are better suited to study data recordings that reflect coarse-grained neural activity, such as EEG (electroencephalogram) or MEG (magnetoencephalogram), which are commonly used in humans and large animals where invasive fine-scale single-neuron resolution studies are not always possible \cite{david2003neural,david2004evaluation,tesler2022mean}. 

A popular phenomenological model is the Wilson-Cowan model, which (in its original form) describes the average firing rates of two homogeneous populations of excitatory and inhibitory neurons \cite{wilson1972excitatory,destexhe2009wilson}. Another example is the Jansen-Rit model \cite{jansen1995electroencephalogram}, which was developed as a model for EEG recordings. Coarse-grained neural-mass models like these have been used to describe a range of phenomena, such as seizure activity observed in epilepsy \cite{meijer2015modeling,shusterman2008baseline}, traveling waves in the visual cortex \cite{wilson2001dynamics,roberts2019metastable}, visual hallucination patterns \cite{ermentrout1979mathematical}, working memory \cite{destexhe2009wilson} and the impact of Alzheimer's disease proteins \cite{ranasinghe2022altered}. While such phenomenological models are insightful, it can be difficult to directly connect them to microscale properties of neurons as they lack sufficient detail \cite{coombes2018next,bick2020understanding}. Instead, bottom-up approaches, wherein neuron models are reduced either exactly or through approximation schemes provide an opportunity to bridge dynamics on microscopic and macroscopic scales \cite{coombes2018next,bick2020understanding,carlu2020mean,breakspear2017dynamic}. A common approach is to start with spiking neuron models that incorporate basic microscopic properties of a neuron (typically a variant of the integrate-and-fire model) and develop a mean-field approximation \cite{renart2003mean,montbrio2015macroscopic,di2019biologically,coombes2018next}. 

Such neural mass models are also key ingredients for modeling approaches used in the Virtual Brain \cite{sanz2015mathematical,sanz2013virtual,roy2014using}, which aims to simulate collective whole-brain dynamics. A key goal of this initiative is to build `digital twins' capturing most relevant features, enabling researchers to test specific hypotheses and intervention strategies for healthcare and personalized medicine \cite{wang2024virtual}. Ongoing efforts focus on applying these models to healthcare applications, including multiple sclerosis \cite{voigt2021digital}, epilepsy \cite{jirsa2023personalised}, ageing \cite{lavanga2023virtual} and Parkinson's disease \cite{angiolelli2024virtual}. A key challenge in building such biophysically principled models is inferring the microscopic parameters that underlie macroscale recordings. Efforts, such as the Human Neocortical Neurosolver (HNN) \cite{neymotin2020human}, are underway to study the circuit and cellular-level mechanisms and origins of measured EEG/MEG signals. A better understanding of these origins could be useful for predictions of neural function and disease, and in probing the effects of stimulation on neural systems \cite{cakan2020biophysically, papadopoulos2020relations}. As our discussion has already reached the regime of large datasets, in the next section we discuss larger brain networks in more detail.

\section{Analyzing large networks}\label{sec:Sec4}

\begin{figure*}
    \centering
    \includegraphics[width=\linewidth]{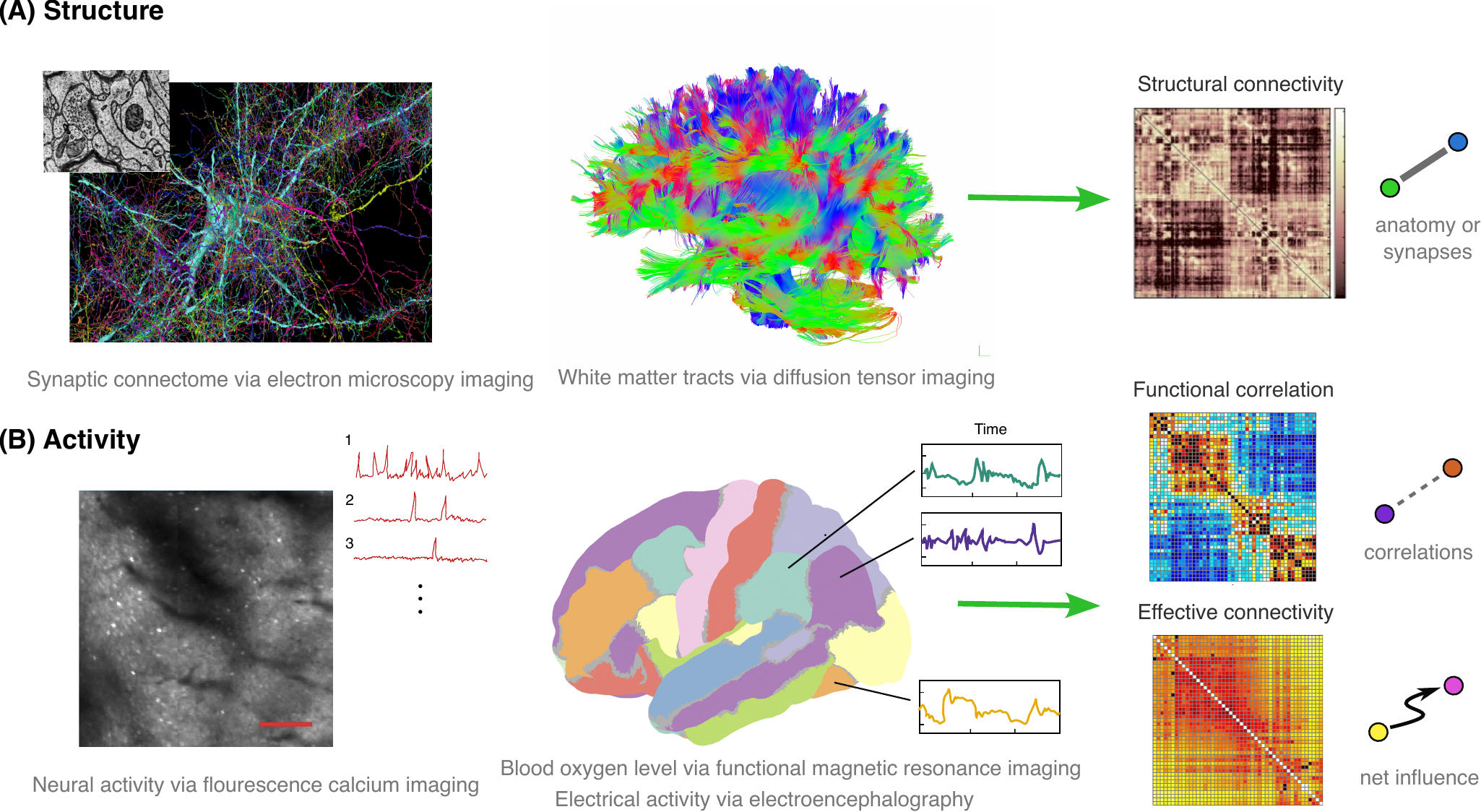}
    \caption{\textbf{Measuring brain network structure and activity.} A range of experimental techniques are used to probe the structure and activity of neural systems at different resolutions. (A) For the fine-scale structure of the brain, electron microscopy imaging can be used to map out neurons and their synaptic connections (left panel, figure reproduced from data in Ref. \cite{shapson2024petavoxel}). To measure the brain's macroscale axonal or white matter organization, diffusion tensor imaging is used (center panel). From these structural data, whether mapping connections at the microscale or at the macroscale, an adjacency matrix can be constructed, with entries weighted appropriately (right panel. Colors represent connection strength). In DTI imaging, nodes represent non-overlapping grey matter volumes while in synaptic connectomes, nodes represent individual neurons. (B) At the microscale, activity of neurons can be measured using fluorescence calcium imaging or electrophysiological methods (left panel, figure reproduced from Ref. \cite{grienberger2022two}). Large-scale brain activity is typically monitored by measuring blood oxygen level variations in different parts of the brain using functional magnetic resonance imaging (fMRI) or by measuring the overall electrical activity using electroencephalography (EEG) (center panel). Similarities between pairs of activity time series can be calculated and represented in a functional correlation matrix (right panel, top). Alternatively, direct or indirect causal measures of the net influence of one region (or neuron) on another can be calculated as effective connectivity (right panel, bottom).}
    \label{fig:Fig2}
\end{figure*}

With experimental advances, we are now obtaining increasingly high-quality and large-scale data on both the structure and activity of neural systems. Tools from complex systems provide a means to represent and analyze this data. Although we primarily discussed networks of interconnected neurons in the previous sections, these tools can be applied to various types of brain network models representing connections or interactions at scales that are coarser or finer. These also include models that do not solely reflect anatomical connections, such as signal propagation maps. Before surveying these methods, we first discuss the kinds of brain networks that researchers typically study. Our goal is to build a high-level understanding of the information these brain networks capture; for more detailed discussion, we refer the reader to recent review articles and books \cite{fornito2016fundamentals, korhonen2021principles, zhang2022quantitative, briggman2012volume}. Very broadly speaking, most brain networks fall into two categories: those that capture relationships of structure and those that capture relationships of activity, as depicted in Fig. \ref{fig:Fig2}.

Advances in electron microscopy and automated image analysis have enabled detailed 3D reconstructions of brain networks at single-neuron resolutions \cite{helmstaedter20083d}. These connectomes map out neurons and the chemical synapses between them. We now have connectomes for a range for species including \emph{C. elegans} \cite{white1986structure}, larval zebrafish \cite{hildebrand2017whole}, \emph{Drosophila} \cite{dorkenwald2023neuronal,court2023virtual,scheffer2020connectome}, and parts of the mouse \cite{microns2021functional} and human cortex \cite{shapson2024petavoxel}. Simultaneously, non-invasive methods have made it possible to study neural circuits in humans and larger animals. Techniques like diffusion MRI tractography and diffusion tensor imaging (DTI) track the diffusion of water molecules to reconstruct white matter tracts connecting distinct brain regions \cite{zhang2022quantitative,fornito2016fundamentals}. Whether mapping synapses linking neurons or white matter tracts connecting brain regions, one can construct structural brain networks that capture how neural elements are wired together. The nodes and edges in these networks are defined according to the method used to quantify them. For networks built from methods that probe the microscale structure, such as electron microscopy, nodes represent individual neurons and edges are generally weighted based on the number of synapses between neuron pairs. For networks built from methods that probe the macroscale structure, such as DTI, nodes represent brain regions and edges are weighted based on the strength and density of the white matter tracts connecting them \cite{fornito2016fundamentals}. Building appropriate representations of structural data is an ongoing challenge, and the best approach depends upon researchers' goals and the data available. It is important to recognize the assumptions and limitations of current methods. For instance, synaptic networks are typically weighted by synapse count, but not all synapses may have the same strength or neurotransmitter type. Similarly, current tractography algorithms used in macroscale studies are unable to resolve directionality of connections \cite{schilling2019limits}. In Section \ref{sec:Sec6}, we elaborate on how more detail can be incorporated in models when available. 

In addition to structural mappings, various experimental techniques are employed to probe brain activity at different scales. At the neuronal scale, calcium imaging and electrophysiological recordings are used to monitor the activity of individual neurons. These methods have enabled simultaneous whole-brain measurements of single-neuron activity in small transparent animals like \emph{C. elegans}, hydra and zebrafish \cite{randi2023neural,yamamoto2020whole,urai2022large}. Efforts are now underway to expand the number of neurons simultaneously tracked in \emph{Drosophila} and cortical neurons in mammals \cite{demas2021high,paulk2022large,jun2017fully}. Neuropixels probes now enable single-spike resolution recordings in large populations of neurons distributed across brain regions in freely moving animals, with the ability to track this activity over the scale of weeks and even months \cite{steinmetz2021neuropixels,paulk2022large,jun2017fully}. The ability to perform such recordings stably over longer timescales opens up exciting opportunities to understand processes such as learning, memory and behavior \cite{steinmetz2021neuropixels}. In humans, researchers use noninvasive methods like electroencephalography (EEG), which records electrical activity through sensors placed on the scalp, and functional magnetic resonance imaging (fMRI), which uses blood oxygen levels in 3D brain images as a proxy for neural activity \cite{fornito2016fundamentals}. When building networks from fMRI or EEG data, researchers measure correlations, coherence, or other statistical dependences between the activity of different regions to quantify functional similarity \cite{fornito2016fundamentals,korhonen2021principles}. We note that correlative approaches, though insightful, do not provide direct measures of causality.

\begin{figure*}
    \centering
    \includegraphics[width=0.8\linewidth]{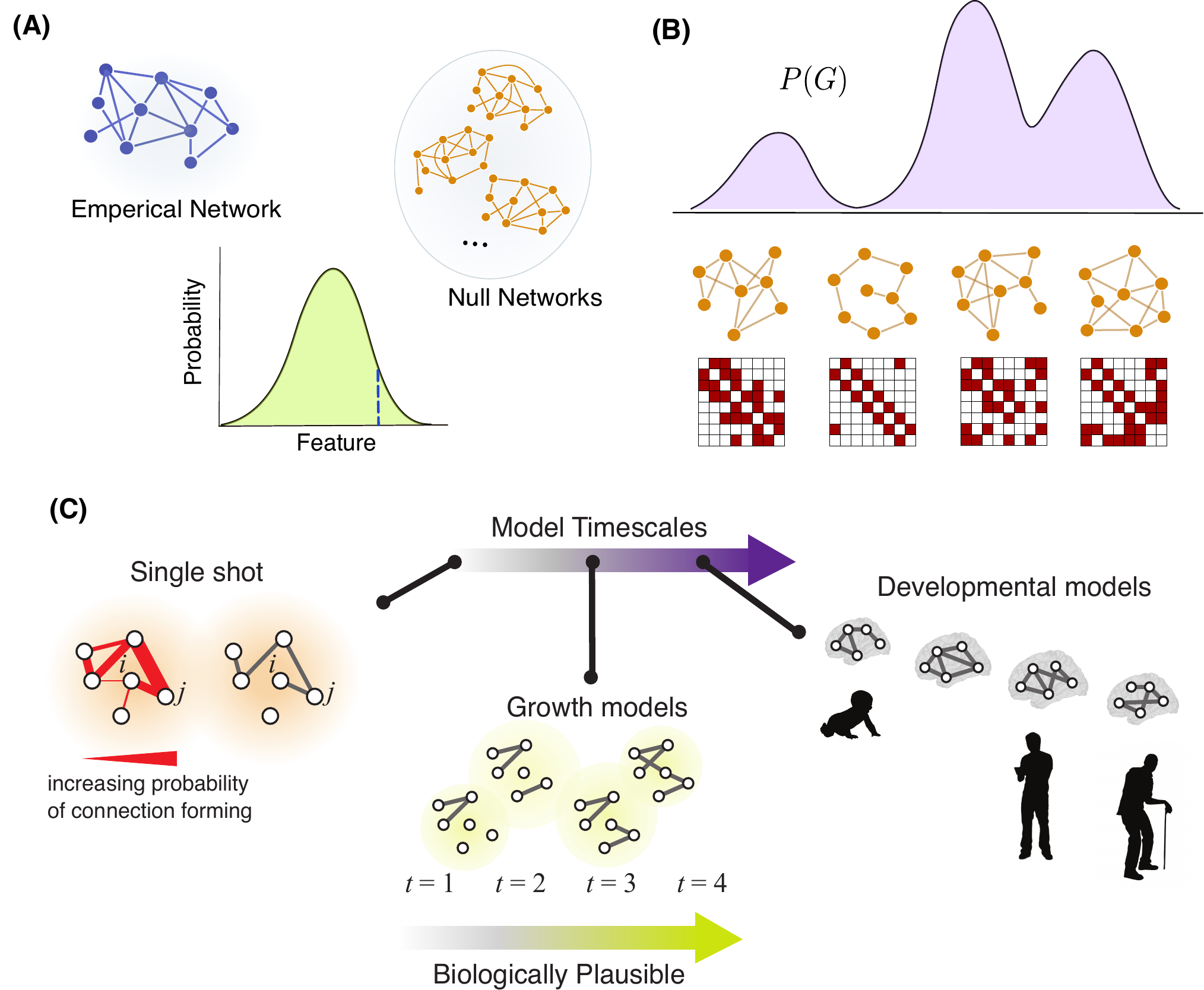}
    \caption{\textbf{Statistical analysis of brain networks.} (A) Comparisons with null networks facilitates a principled assessment of how significant an empirically observed property is. To determine whether a feature of the empirical network is statistically unlikely, one can generate an ensemble of null networks. These null networks preserve specific desired properties of the empirical network, forming the null hypothesis, but are otherwise random. The feature of interest can then be calculated for each null network to obtain a probability distribution of the feature under the null hypothesis. (B) The exponential random graph model enables the generation of ensembles from the ground up that constrain certain properties, producing a probability distribution in the space of all possible graphs. Such an ensemble can then be used to test whether the constrained properties are explanatory of an observed network's structure. These ensembles can also function as null networks. (C) More broadly, generative modelling is a framework for constructing synthetic networks based on a set of rules. The space of generative models can span different timescales. In single shot models, connection probabilities are defined early on, and the network is generated in a single step. More biologically realistic models operate on timescales closer to that of development and can be used to model network growth and development. Intermediate models lie between these two extremes, where the network grows over arbitrary timescales that do not have biological interpretation. Figure in panel (C) reproduced from Ref. \cite{betzel2017generative}.}
    \label{fig:Fig3}
\end{figure*}

From these datasets, tools developed in physics and mathematics enable researchers to uncover patterns of organization that may not otherwise be immediately apparent. Across different organisms and spatial scales, such efforts have identified properties common in nervous systems. These include small-world properties \cite{bassett2017small}, heavy-tailed edge weight and degree distributions, modularity \cite{hilgetag2000anatomical,sporns2016modular}, and the existence of hubs \cite{van2011rich,towlson2013rich}, among others. The presence of these features across vastly different species suggests that there might be general organizing principles at play and that these features might support specific functions, which we explore in detail in Section \ref{sec:Sec5}. While network science provides an array of metrics to quantify organization, we emphasize that the goal has never been to be merely descriptive. Instead, we seek insight into how these properties arise and their implications for neural function \cite{bassett2018nature}. 

When examining a new network and its features, comparisons with null models enable researchers to benchmark the significance of these features \cite{vavsa2022null} (Fig. \ref{fig:Fig3}A). Null models are ensembles of networks that preserve certain selected features of a network but are otherwise random. In the best scenario, null models can disentangle a given network feature from other network properties, though this may not always be possible. The selection of graphs for sampling null models requires careful consideration. For example, the Erd\H{o}s--R\'enyi graph model, which places edges between nodes with equal probability, often does not reflect processes or constraints relevant to the brain. Conversely, overly stringent null models that constrain the space of graphs too much can limit how generalizable the results are. Constraints inherent to the empirical network, such as the presence of self-loops or multi-edges, can be usefully reflected in the null graph space, as these can directly impact conclusions \cite{fosdick2018configuring}. In practice, null networks are typically obtained by either rewiring the original network according to specific rules or building them from the ground up using generative models \cite{vavsa2022null}. An example of a rewiring model is the degree-preserving configuration model, which involves rewiring the graph using double-edge swaps among edges \cite{fosdick2018configuring}. This model retains the exact degree sequence, preserving features like hubs and heavy-tailed degree distributions that are typical of brain networks, allowing researchers to assess how the network's degree sequence influences features of interest. Rewired null models can also be constructed to make the edge swaps based on other constraints, such as edge weight or path lengths. 

In contrast to rewiring models, generative models build networks from scratch using specific rules, such as homophilic attachment or minimizing the total edge length \cite{betzel2017generative,vavsa2022null}. Generative modeling is also insightful in examining design principles that shape network features. Since brain networks are an outcome of biological processes that are inherently noisy, it is often useful to view neural systems probabilistically. That is, not as a fixed network, but as a distribution over a space of networks informed by biologically-motivated design principles. Exponential random graph models (ERGMs), provide a means to identify such local design principles behind real networks \cite{dichio2023statistical, simpson2011exponential}. A part of the family of maximum entropy models \cite{jaynes1957information}, the exponential random graph probability distribution reproduces certain selected features, but is otherwise maximally random (Fig. \ref{fig:Fig3}B). While all our discussion so far has focused on static networks, we know that brain networks---both structural and functional---are dynamic. Exponential random graphs can be appended with this information by creating a time series and estimating parameters for this sequence. These are referred to as temporal exponential random graph models \cite{dichio2023statistical}. 
 
Methods based on the maximum entropy principle are also used to construct models of neural activity, complementing the biophysical modeling approaches discussed in Section \ref{sec:Sec3}. Instead of starting with a biophysical model, this approach involves constructing probability distributions over the space of network states that match selected observed properties of the system while otherwise remaining as random as possible \cite{jaynes1957information,schneidman2006weak}. Initially, pairwise maximum entropy models were developed by constraining over the average activity and correlations between pairs of elements \cite{schneidman2006weak,ashourvan2021pairwise}. Although this method effectively captured correlations among tens of neurons, it struggled to capture correlations among hundreds of neurons sampled across more distant subgroups \cite{meshulam2021successes,ganmor2011sparse}, hinting at the role of higher-order correlations for such samples. While higher-order correlations can be incorporated manually to improve these models \cite{ganmor2011sparse}, a more scalable approach for larger datasets is based on random projections \cite{maoz2020learning}. The philosophy behind random projections is also rooted in statistical mechanics: higher-order interactions are captured in the form of random projections, and the structure of this randomness is constrained using data. A less explored class of models are maximum caliber models, which generalize the maximum entropy principle to account for dynamical correlations \cite{presse2013principles}. Instead of constraining correlations at the same time point, these models constrain correlations between current and future time points.

While concepts stemming from the network theory literature have provided a valuable starting point for gaining insight into data from neural systems, they need to be adapted or generalized to quantify structures coming out of the latest neuroscience experiments and for dynamics relevant to neuroscience \cite{curto2019relating}. Future progress in this field will be shaped by how effectively we can bridge data and models. For example, it is known that spatial structure plays an important role at different scales and brain networks are spatially embedded. These constraints can be usefully captured in null models of brain networks \cite{stiso2018spatial,salova2024combined}. We discuss ways in which current models can be expanded to derive meaningful insight from experimental data in Section \ref{sec:Sec6}.

\section{Design principles and functional affordances} \label{sec:Sec5}

We have discussed so far how advances in experimental research and mathematical tools have enabled systematic analysis of neural systems. Despite the inherent stochastic and noisy nature of biological processes, neural systems display remarkable structure and organization across various spatial and temporal scales. Many of these features are observed across vastly different organisms, suggesting general design principles at play \cite{van2016comparative}. Alongside these common features though, distinctions across organisms and sensory modalities within an organism could reflect specific adaptations and different constraints. Indeed, the brain must perform extremely complex and diverse processes under stringent resource constraints---be it space, material, energy or time---in a manner that is fast and adaptable. This fact raises important questions: What biological design principles shape the organization of neural systems across spatial scales and in different functional domains? How does the observed organization enhance information processing given various constraints? Recent experimental and modelling advances can be used to infer potential design principles and test hypotheses out across various organisms and developmental stages. These principles can illuminate how neural systems achieve high levels of performance and how their structure supports function. Additionally, such principles could offer insights into how perturbations might affect neural circuit organization and function, with implications for behavior and dysfunction. In this section, we discuss various design principles thought to shape the structure of neural circuits. The central questions are twofold: Why are neural circuits structured the way they are, and what biologically relevant functions or advantages do these structures provide? While determining if these principles are indeed \emph{causal} may require further experimental characterization, they nevertheless provide valuable general insights.

\begin{figure}
    \centering
    \includegraphics[width=\linewidth]{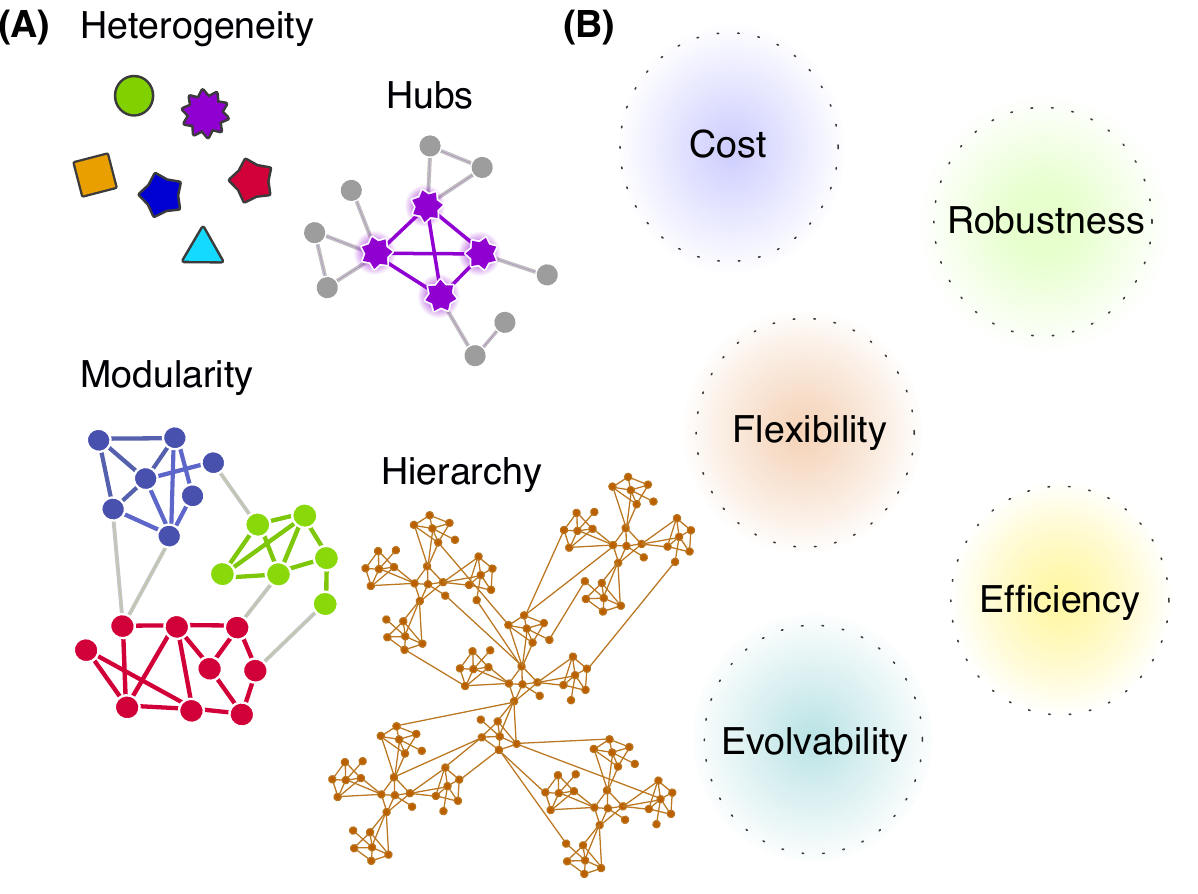}
    \caption{\textbf{Organizing principles that shape the structure and working of neural circuits at various scales.} (A) Network motifs commonly observed in neural systems show immense heterogeneity in both their components and interaction patterns. Many large-scale brain networks possess community structure, displaying modularity at several scales organized hierarchically. Large-scale brain networks also feature high-degree hubs that form a densely connected core. (B) These observed patterns could be shaped by various biologically-motivated design principles. Neural systems balance managing energetic and material costs, being robust to noise and external perturbations, and remaining flexible enough to adapt to new environments. While optimizing these constraints, their functioning must also be efficient. }
    % Modify figure for hierarchy
    \label{fig:Fig4}
\end{figure}

Several factors influence the architecture and functioning of brain networks, including the need to manage metabolic and material costs \cite{bullmore2012economy}, ensure efficient information transmission \cite{bullmore2012economy,van2012high}, maintain robustness to noise or injury, and adapt to process new information (Figure \ref{fig:Fig4}). However, prior to discussing these factors, we first emphasize that brain networks are inherently spatial. Circuits in the brain are physically embedded in 3D space within a fixed enclosure and the neurons themselves have intricate spatial structure. This spatial nature constrains brain growth, development, and function \cite{stiso2018spatial}. It is important to consider this fact when applying tools from network science, as certain models do not account for spatial information. One of the earliest observed features of (larger) neural systems is their segregation into specialized anatomical modules that interact with one another. Within each module, there is further specialization; as concrete examples, in large mammals, the visual system has functional areas that process different visual features, while the motor system has regions that control different parts of the body. Even within these segregated regions, there is enormous diversity in neuron types, each performing specialized tasks. Thus, at every scale---from the entire brain down to single cells---the brain is composed of heterogeneous components that are arranged in modules which interact hierarchically, simultaneously segregating and integrating information to produce high-level responses. Neural heterogeneity is thought to shape computational functions differently in different cell types, providing a means to control properties of neural circuits \cite{gast2024neural}. Several studies have examined the modular and hierarchical structure in both structural and functional networks across organisms \cite{bassett2008hierarchical,meunier2010modular,hilgetag2000anatomical}. This modular structure reflects segregation and functional specialization, which can also make the network robust to perturbations, either internal (in the form of noise or genetic variation) or external (in the form of injury) \cite{kaiser2011evolution}.

Another feature of large-scale brain networks is their small-world architecture \cite{bassett2017small, liao2017small, watts1998collective}. That is, in addition to high clustering in tightly connected modules, the brain has short characteristic path lengths. This architecture supports the idea that the brain balances segregation and integration, with modules for local information processing and efficient routes for long-range communication. More generally, brain network communication, the study of how information is transmitted throughout the brain in an efficient and biologically plausible manner, is a rich area of research \cite{seguin2023brain}. Indeed, the brain has to process and communicate immense amounts of information under strict time, energetic, and material costs. Physical connections in the brain come with a cost of building and maintenance, a factor which becomes important particularly for larger brains. However, the presence of many of the complex properties discussed earlier---including the high efficiency of information transfer between regions that are anatomically distant---suggests that the brain does not strictly minimize wiring costs. Rather, the pressure for minimizing wiring competes with pressures for efficient communication \cite{kaiser2006nonoptimal,bassett2010efficient,budd2012communication}, and the brain negotiates a trade-off between these factors. In addition to the modular and small-world structure, large-scale brain networks are heterogeneous, featuring nodes that form a rich core of densely connected `hubs' \cite{van2011rich}. These hubs are thought to play important roles in a range of functions, serving as broadcasters of information. Further, these hubs also reduce the overall average path length across the network, supporting integration of information.

From a metabolic standpoint, the brain is one of the most expensive systems within an organism: the human brain is 2 \% of the body weight but 20 \% of its metabolic load \cite{sterling2015principles, harris2012synaptic}. How the brain manages its energy budget among the processes carried out by its units is an area of active research. Some findings suggest that the cost involved in communication is more substantial than the cost involved in computation \cite{levy2021communication}. As discussed earlier, this possibility has implications for the structure of neural systems at each scale. At the microscale, it has been observed that the distribution of mitochondria and the thickness of axons correlates with typical firing rates. At the macroscale, the features discussed in previous paragraphs could contribute to metabolically efficient communication. Thus, such heterogeneity could in part support the brain's efficiency \cite{balasubramanian2015heterogeneity}. Turning from communication to computation, we note that despite the high metabolic load of the brain, the cost of computation is still cheaper than artificial machines. How does the brain achieve this efficiency in computation? One idea is that this efficiency exists in part because circuits are adapted to the structure of the world, and have the capacity to adapt further in the future \cite{palmer2015predictive,wang2021maximally,price2022efficient}. Additionally, neural codes must also be able to overcome noise while representing information. One potential way to do this is by choosing codes that are sufficiently far apart in the space of patterns \cite{curto2013combinatorial}. Finally, the consumption of energy to perform function is inherently a non-equilibrium process and the framework of stochastic thermodynamics may be fruitful in examining some of these questions. 

\section{Current frontiers}\label{sec:Sec6}

\begin{figure*}
    \centering
    \includegraphics[width=\linewidth]{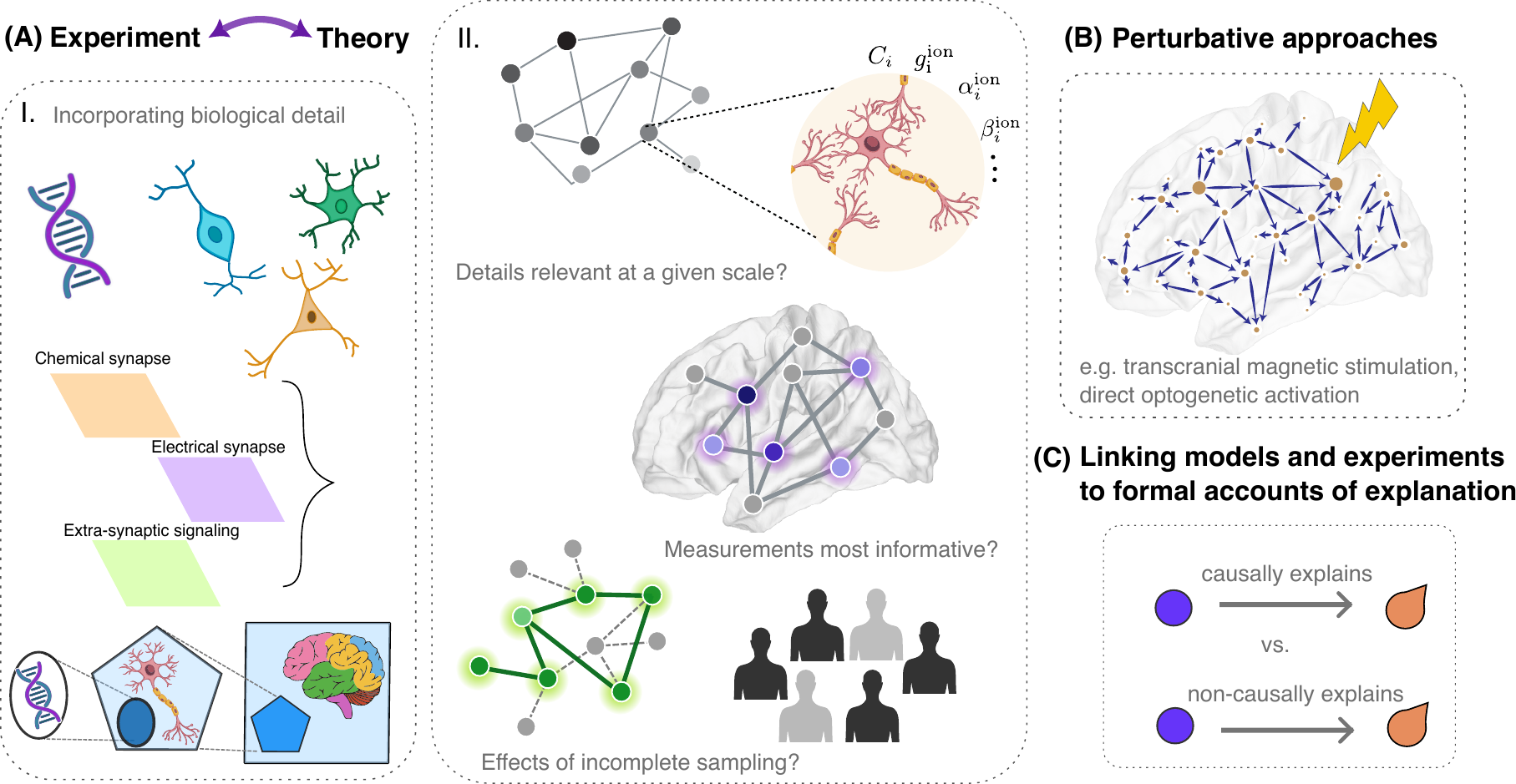}
    \caption{\textbf{Ongoing and emerging future directions.} (A) Effectively bridging experimental data and models is an important area of ongoing research. These involve: (I) Expanding theoretical tools and models to incorporate biological detail to make sense of connectomes. These expanded models also motivate efforts to bridge between different scales of description. (II) Using theoretical tools to guide experimental efforts and measurements. This could involve identifying which measurements are most information of behaviors of interest at a given scale and examining the effects of sparse measurements. (B) Perturbative techniques provide new opportunities to improve our understanding of neural systems. (C) Linking the contributions of modeling and experiments to distinct kinds of explanations can inform the design of effective experimental and theoretical approaches. }
    \label{fig:Fig5}
\end{figure*}

We now discuss exciting opportunities for research aimed at improving our understanding of neural systems in light of recent advances (Figure \ref{fig:Fig5}). Broadly speaking, we classify these directions into four key themes. The first theme involves expanding theory and modeling to keep up with data coming in from large-scale experiments. The second relates to how theory can feed back into experiments and guide measurements. Together, these two themes contribute to strengthening the loop between theory and experiment. The third explores how methods for perturbing neural systems open new ways of understanding the brain. Finally, the fourth theme emphasizes linking modeling and experiments with formal accounts of explanation and causation to contextualize their contributions.
\subsection{Expanding models to make sense of connectomes}
As discussed in Sections III and IV, tools from network theory have been applied to connectomes at various scales---from macroscale connectivity between brain regions \cite{van2011rich,li2013mapping,elam2021human} to microscale synaptic connectivity between neurons \cite{varshney2011structural,lin2023network,lynn2024heavy}---to uncover patterns of organization. These approaches also provide a quantitative means to compare the organization and working of neural systems across individuals within the same specie, across different developmental stages \cite{witvliet2021connectomes}, and across distinct species \cite{van2016comparative,brynildsen2023network}. With advances in high throughput electron microscopy and reconstruction methods, the microscale structure of neural systems is becoming increasingly accessible, even in larger organisms \cite{dorkenwald2023neuronal,shapson2024petavoxel,abbott2020mind}. Researchers have employed graph-theoretical tools to study these neuron-level connectome datasets, such as chemical synapses and gap junctions in \emph{C. elegans} \cite{varshney2011structural} and the whole-brain connectome of \emph{Drosophila} \cite{lin2023network}. These studies are identifying motifs crucial for computation and neural function as well as uncovering general mechanisms that underlie observed network patterns.

At the same time, existing mathematical tools can be expanded to reflect biologically relevant information when necessary (Figure \ref{fig:Fig5} (A), Panel I). For example, attributes such as cell types, neurotransmitter receptor profiles, gene expression, tendency to activation during certain tasks can be crucial to build a complete picture of neural organization, activity, and function. One approach to improve the biological descriptiveness of existing models is to overlay these attributes onto nodes and edges as `metadata', creating annotated networks \cite{newman2016structure, bazinet2023towards}. This method can bridge abstract models with important microscale features, opening up new opportunities for discovery. It enables researchers to explore sophisticated questions, such as how observed patterns of structural connectivity relate to micro-scale biological attributes \cite{bazinet2023assortative,bazinet2023towards}. In addition to biological attributes, traditional graph theory methods can also be usefully adapted to reflect the spatial nature of brain networks. Incorporating both spatial and biological information into null models can enhance their relevance and utility. Although developing null models can involve tricky choices, improving them to account for the spatial and biological aspects of networks will improve our ability to address sophisticated research questions.

Other useful extensions of network representations include multilayer \cite{kivela2014multilayer,vaiana2020multilayer} and higher-order networks \cite{battiston2020networks,bick2023higher}. Multilayer networks provide a framework to study different kinds of interaction within neural systems and to integrate multiscale data \cite{vaiana2020multilayer,presigny2022colloquium}. For instance, the connectome of C. elegans can be modeled as a multilayer network to capture different modes of communication, including synapses, gap junctions, monoamines, and neuropeptides \cite{bentley2016multilayer}. Multilayer networks can also capture temporal information, allowing researchers to examine how networks evolve over time  \cite{holme2012temporal,thompson2017static,bassett2013task}. Capturing temporal aspects of brain activity and structure, including modeling the relationship between dynamics \emph{on} networks and dynamics \emph{of} networks \cite{berner2023adaptive} is a crucial area for current and future efforts. Although these approaches offer richer descriptions of neural systems, they require appropriate experimental data and careful consideration to ensure suitability for specific research questions.

\subsection{Using theoretical analysis to guide experiments and measurements}

As highlighted throughout this review, applying theory and modeling to large-scale recordings offers valuable insights into neural systems. But how exactly can these models then inform experimental efforts? Biophysical models of large-scale brain activity often involve on the order of hundred parameters. However, previous research has demonstrated that a range of multi-parameter models across physics and biology exhibit `sloppiness', meaning that their behavior is primarily influenced by a few combinations of parameters, while the rest are relatively unimportant for predictions \cite{transtrum2015perspective,gutenkunst2007universally,machta2013parameter}. Applying this approach to models of large-scale neural activity could illustrate how sensitive these models are to different parameters (Figure \ref{fig:Fig5} (A), Panel II). Identifying which parameters are critical at a given scale could then help guide experimental efforts to measure these parameters more reliably or even inform the selection of the most appropriate models. 

As the community advances towards tracking thousands of simultaneously recorded neurons across multiple brain areas, the size of datasets is expanding rapidly. This growth makes it increasingly impractical to monitor all correlations in neural activity, highlighting the need for reliable models and analysis techniques that can handle these larger datasets. Fortunately, recent research suggests that while the space of possible states of a network is vast, neural activity often remains constrained to a lower-dimensional manifold \cite{cunningham2014dimensionality,gallego2017neural,gao2015simplicity}. Rather than relying on individual neural activity, neural function may be better understood through population-wide activity patterns, known as neural modes. This has led to a range of studies aimed at manifold identification.
In parallel with efforts to study an increasing number of brain regions, another approach could be to use theoretical insights to identify which populations or regions provide the most unique (or least redundant) information about neural activity (Figure \ref{fig:Fig5} (A), Panel II). This strategy could be especially valuable in human neuroscience, where it can guide more effective probe placement. By optimizing probe placement, researchers may be able to reduce the number of required electrodes, thereby decreasing health risks.  

Finally, we point out that most studies of larger organisms are based on incomplete sampling, either by recording only a subset of regions \cite{conrad2020sensitivity} or by examining only a small population \cite{marek2022reproducible}. Though recent strides in population-level recordings allow for simultaneous recording of up to tens of thousands of neurons, this still is a small fraction of circuits in larger animals. Sampling from a subset could create errors in models of collective behavior. This issue calls for quantifying the kinds of errors that partial measurements could create, and for methods to infer models from such partial data reliably (Figure \ref{fig:Fig5} (A), Panel II). A deeper understanding of the effects of partial measurements could in turn guide experimental efforts and how to allocate resources, potentially identifying regions that could benefit from more sampling and regions which do not require as much. Additionally, systematically examining noise in experimental measurements that could bias measurements is key to maximizing the impact of experimental methods \cite{urai2022large}. 

\subsection{Perturbation approaches might open new ways of modeling and understanding neural systems.}

Anatomical maps of the brain provide some information regarding circuit structure. However in certain cases, the exact nature of the inputs and outputs including their signs, strengths, and timescales cannot always be established. Hence, these maps do not capture the full picture behind neural function. On the other hand, correlative approaches, which attempt to infer function based on activity do not measure causality and might not capture the detailed mechanism behind observed activity. The recent few years has seen improvement in methods to directly perturb neural elements---be it brain regions in humans or even single neurons in smaller animals---and measure the propagation of activity among other elements (Figure \ref{fig:Fig5} (B)). These perturbative approaches can directly measure signal propagation, capturing all kinds of interactions and processes that might be present. Such approaches are more suited to studying network-level phenomena and are also causal. By using simultaneous stimulation and recording, the interactions or network can be inferred in a principled manner \cite{lepage2013inferring}. In humans, methods like Transcranial Magnetic Stimulation (TMS) are increasingly being used for such perturbations \cite{walsh2000transcranial}.  For smaller organisms, it is possible to stimulate even single neurons with light or ablate groups of neurons \cite{randi2023neural}. These smaller organisms can thus serve as a test bed for simulating the behavior of an entire nervous system. These perturbative approaches have the potential to open new avenues for neural system modeling and understanding, offering exciting opportunities to test hypotheses derived from models.

% Include perturbations at the fine-scale: to neurotransmitters or gene expression?

\subsection{Linking theory and modelling to different kinds of explanations}

As efforts progress on the experimental and theoretical fronts, it is also worth discussing the kinds of explanations that can be sought and gained (Figure \ref{fig:Fig5} (C)). Explanations come in a variety of kinds, and can be distinguished in multiple ways. Perhaps the most commonly raised distinction is between causal and non-causal explanations \cite{sep-scientific-explanation}. Causal explanations are often sought to explain why or how a phenomena occurs, while non-causal explanations are often sought to explain what a phenomenon is. However, it is also possible to identify non-causal explanations to answer why and how questions. An prime example is a mathematical explanation, wherein an equation is used to explain a phenomenon such as a process, behavior, or dynamics \cite{sep-mathematics-explanation}. 

Here in this review we have focused on approaches from statistical physics, information theory, and network science that can be used to explain principles of brain network architecture and function. Do these approaches provide causal explanations, non-causal explanations, both, or neither? When our methods focus on mathematical descriptions (such as, for example, the maximum entropy model), we may be excavating non-causal mathematical explanations. When our methods focus on network topology, we may be excavating structural explanations. In some cases, structural explanations can be causal, as has been investigated in regard to social phenomenon \cite{ross2024what} and topological factors \cite{ross2021distinguishing}, whereas in other cases, structural explanations can be non-causal \cite{kostic2020general}. In addition to causal structural explanations, the approaches canvassed in this review also have potential to provide other causal factors, including pathways \cite{ross2021causal}, mechanisms \cite{ross2024causation}, cascades, triggering causes \cite{Dretske2010-DRETAS}, and causal constraints. 

The accessibility of this causal diversity motivates a renewed examination of how we as scholars approach scientific problem solving \cite{ross2021causes}. Detecting \emph{a} cause is less satisfying than determining \emph{what type of} cause is involved and why. How might we design experiments, models, and theory to distinguish types of causation in brain network function? How might distinctions between causal relationships influence how we engage in causal reasoning or use causal knowledge to control the functional outcomes of neural systems? Engaging these and related questions is important as we aim to shape future research and the training of young scholars interested in understanding the brain. 

\section{Outlook}

The brain is an astounding feat of natural engineering whose structure and function is becoming increasingly accessible to the tools of modern science. As experimental techniques continue to mature, physicists generally and biophysicists specifically have contributed markedly to the development and application of models and theory. Here in this review, we have canvassed those contributions and associated recent advances in the use of statistical mechanics, information theory, and network science with the goal of identifying principles of brain network architecture and function. Spanning from models of single neurons and small groups to large-scale ensembles and networks, the efforts examined here have proven useful in isolating design principles of neural systems and the functional affordances thereof. As these efforts continue to produce insights in the coming years, we envision several critical frontiers. These include expanding models to make sense of connectomes, using theoretical analysis to guide experiments and measurements, and using perturbative approaches to provide new kinds of understanding. Throughout these (and other) expansions, we think it will be critical to link the contributions of modeling and theory to distinct sorts of explanation. By pursuing both causal and non-causal explanations, and by detecting specific types of causes, the field of neurophysics will be poised to better design effective and specific experimental, computational, and theoretical approaches, better engage in causal reasoning, and better use causal knowledge to control the functional outcomes of neural systems.

\section{Acknowledgements}

We thank L. Papadopoulos, C. W. Lynn, A. Winn, C. G. Alexandersen, J. K. Brynildsen, I. Stallworthy and S. Patankar for helpful comments on earlier drafts.
\bibliography{MAIN_review_draft}
\end{document}